\documentclass{article}
\usepackage{spconf,amsmath,graphicx}
\usepackage{hyperref}
\usepackage[capitalise]{cleveref}
\usepackage{amsmath,amssymb,amsfonts}
\usepackage{xcolor}
\usepackage{subcaption}
\usepackage{booktabs}
\usepackage{orcidlink}

\def\minwrt[#1]{\underset{#1}{\mathrm{minimize }}}
\def\mminwrt[#1]{\underset{#1}{\mathrm{min }}}

\newcommand{\RR}{\mathbb{R}}

\newcommand{\signedwass}{\mathbb{W}}
\newcommand{\posmeas}{{\mathcal{M}_+}}
\newcommand{\signedmeas}{{\mathcal{M}}}
\newcommand{\edc}{{\varepsilon}}
\newcommand{\Ldist}{{\mathcal{L}}}

\title{Signed Wasserstein distance for room impulse response comparison}

\name{Linda Fabiani$^{\star}$ \orcidlink{0009-0007-2088-6290}, Sebastian J. Schlecht$^{\dagger}$ \orcidlink{0000-0001-8858-4642} \thanks{This work was supported in part by Finland's Ministry of Education and Culture’s Doctoral Education Pilot under Decision No. VN/3137/2024-OKM-6 (The Finnish Doctoral Program Network in Artificial Intelligence, AI-DOC), as well as Nokia University Donation 2024. \\SJS were supported through the joint German Academic Exchange Service (DAAD) and Research Council of Finland (RCF) Project (57763119) "Immersive Augmented Acoustics (IAA)" }, Filip Elvander$^{\star}$ \orcidlink{0000-0003-1857-2173}}
\address{$^{\star}$ Dept. of Information and Communications Engineering, Aalto University, Finland \\
      $^{\dagger}$ Multimedia Comms. \& Signal Process., Friedrich-Alexander-Universität Erlangen-Nürnberg, Germany}

\begin{document}
\fontsize{9.68}{11.5}\selectfont
%
\maketitle
\begin{abstract}
The nature of room impulse response signals is often adverse to a clear evaluation of similarity: they are sparse, rapidly varying, signed signals resulting from the complex interaction of room reflections and phase relationships. Therefore, conventional point-wise metrics may be sensitive to small temporal shifts and provide poor loss surfaces for optimization. In this paper, we propose an optimal transport-based distance for RIR similarity that explicitly accounts for the structure of impulse responses, inspired by the signed Wasserstein distance formulation. Optimal transport provides a natural framework for this problem, by comparing signals based on how their underlying distributions are displaced rather than requiring a one-to-one correspondence. We evaluate the proposed distance against commonly used audio baseline metrics through a series of room-parameter perturbation studies. The results demonstrate the advantage of using the signed Wasserstein metric for geometric evaluation tasks and show a promising performance for room parameter optimization within a differentiable Image-Source model. 
\end{abstract}
\section{Introduction}
\label{sec:intro}

Evaluating the similarity of signals is a common but far from trivial problem in audio signal processing. For room impulse responses (RIRs), defining a meaningful similarity measure is an equally challenging task. Room impulse responses describe the sound propagation in an enclosed space and exhibit a highly structured temporal behaviour, through a combination of sparse early reflection peaks and a reverberant tail segment \cite{fiftyyearsreverb}.
As a result, RIR characterization relies on these time-localized, signed signals, where even a small variation of room parameters can lead to a large deviation in identifying the corresponding target acoustic environment.
This problem arises in a number of applications, ranging from room impulse response estimation and interpolation \cite{ pallewela2025room, sundstrom2024optimal}, as well as optimization of parametric reverberation models like feedback delay networks \cite{dal2024similarity, mezza2024dfdn}. In these settings, the choice of similarity measure is critical to obtain a loss landscape favorable for optimization. Conventional point-wise objectives, such as  $L_p$-losses can become poorly informative when corresponding reflections are  misaligned in time, causing a large sample-wise error at the global  scale  \cite{Fabiani_Schlecht_Elvander_2024, Sambridge_Jackson_Valentine_2022}. Spectral losses are also used to evaluate distances in terms of frequency content, but additionally suffer from their own limitations due to the choice of time-frequency resolution. \cite{turian2020im, schwar2023multi}. 

\begin{figure}[t]
    \includegraphics[width=\columnwidth, trim=00 0 30 30, clip]{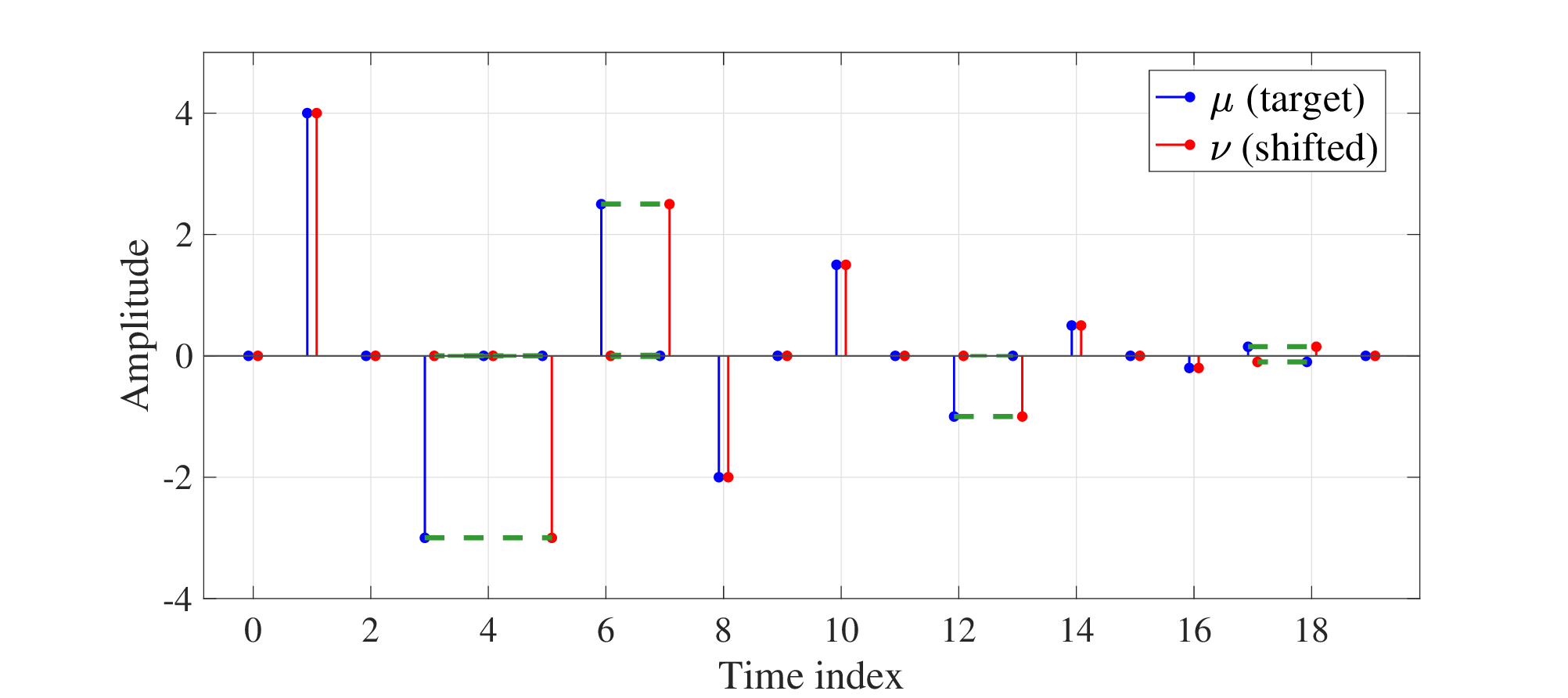}
    \caption{Optimal transport between signed distributions, where the optimal transport plan identifies the most efficient shifts between corresponding positive and negative peaks.}
    \label{fig:illustration}
\end{figure}

Optimal Transport (OT) provides a promising solution to this problem, thanks to its ability to compare distributions while accounting for the geometric displacement of the signals. The use of optimal transport was shown to be beneficial for audio comparison tasks, as well as effective loss function in deep learning frameworks \cite{Fabiani_Schlecht_Elvander_2024, Torres_OT, flamary2016}.
Interestingly, we can identify a similar connection in a common representation in room acoustics. Under appropriate normalization, the $L_1$ distance between Energy Decay Curves (EDCs) admits an exact interpretation as Wasserstein-1 distance of their temporal energy distributions. To our knowledge this equivalence has not been made explicit before, and it shows that OT is already closely related to well established energy-based distances for RIRs. However, the EDC representation discards the information tied to the signed structure of the raw RIRs, limiting the available signal characteristics employed in the similarity evaluation. 

Motivated by this, we propose an OT-based similarity metric that operates directly on signed RIR waveforms. A conventional Wasserstein-1 distance does not directly apply here, since it assumes non-negative measures, while RIRs contain both positive and negative peaks. Thus, we build on the \textit{signed} Wasserstein formulation, which lets us define a transport problem directly on the raw RIR signals. Our main contributions in this work are:
\begin{itemize}
\item We show that, under equal-total-energy normalization, the $L_1$ distance between EDCs is equivalent to the Wasserstein-1 distance between their corresponding temporal energy distributions. This establishes an exact optimal-transport interpretation of a common room-acoustics distance.

\item We introduce a signed Wasserstein-1 distance for raw, signed RIR signals. The formulation preserves the non-local comparison of the EDC-based distance while retaining the sign structure of the RIR, avoiding the treatment of interference effects as purely additive energy.

\item We evaluate the proposed metric through controlled perturbation experiments involving source-position shifts and room-dimension scaling, as well as room-parameter optimization. The results show that the signed Wasserstein metric provides smooth and informative loss landscapes for geometric perturbations and achieves improved room-dimension accuracy compared with the considered baselines. 

\end{itemize}

\section{Optimal Transport}

Optimal transport is a popular mathematical tool used to compare distributions, paying particular attention to their geometric domain. The distance is found by solving an optimization problem, looking for the most efficient way to transport one distribution to another which minimizes a defined cost of transport, as shown in \cref{fig:illustration}. Of particular interest is the special class of \textit{Wasserstein} distances, where the cost of transport corresponds to powers of the Euclidean distance.
For the purposes of this paper, let $\posmeas(\RR)$ be the set of non-negative measures on $\RR$ and let $\mu, \nu \in \posmeas(\RR)$ with the same total mass, i.e., $\mu(\RR) = \nu(\RR)$. Then, the Wasserstein-$p$ distance, for $p \geq 1$, is defined as \cite{villani2008optimal}
\begin{align*}
    W_p(\mu,\nu) = \left( \mminwrt[m \in \Pi(\mu,\nu)] \int_{\RR\times\RR} |x-y|^pdm(x,y) \right)^{1/p}
\end{align*}
where $\Pi(\mu,\nu) = \{ m\in\posmeas(\RR\times\RR)\mid \int dm(\cdot,y) = \mu\:,\: \int dm(x,\cdot) = \nu  \}$ is the set of valid couplings between $\mu$ and $\nu$.
In the one-dimensional case, the Wasserstein-1 distance has a particularly simple expression, namely
\begin{align}\label{eq:wasserstein1}
    W_1(\mu,\nu) = \int_{\RR} | F_\mu(t) - F_\nu(t)|dt,
\end{align}
where $F_\mu$ and $F_\nu$ are the cumulative distribution functions (CDFs) of $\mu$ and $\nu$, respectively. The Wasserstein-1 distance and its extension to signed measures will be central in the rest of the paper.

\section{Room Acoustic Model}

We model RIRs as elements of the space of signed measures on the real line, $\signedmeas(\RR)$. We will view these as atomic measures, i.e., of the form
\begin{align}\label{eq:RIR_atomic}
    h = \sum_{n} \alpha_n \delta_{t_n},
\end{align}
where $\alpha_n \in \RR$ and $\delta_{t_n} = \delta(t-t_n)$ is a delayed Dirac delta. The amplitudes $\alpha_n$ and arrival times of reflections $t_n$ are functions of the room geometry, the source and receiver positions, and other physical constraints, such as the speed of sound. For an RIR $h$, we denote by $h^2$ the non-negative measure
\begin{align*}
    h^2 = \sum_{n} \alpha_n^2 \delta_{t_n}, 
\end{align*}
representing the instantaneous energy of the RIR. 

It must be noted that, under a small perturbation of any of these underlying parameters, both $\alpha_n$ and $t_n$ are perturbed. Variations of the amplitudes $\alpha_n$ are well captured by point-wise metrics, as these operate directly on the values compared at fixed positions. Perturbations of the arrival times $t_n$, however, are not, since it is impossible for such metrics to quantify the magnitude of temporal shifts. This motivates a comparison framework that is explicitly sensitive to such temporal displacements.

\section{Proposed Metric}
In this work, we analyze two similarity metrics for room impulse responses from an optimal transport perspective. First, we show that a commonly used EDC-based distance can be interpreted as the Wasserstein-1 distance between temporal energy distributions. This provides an interesting theoretical connection between room acoustics and the mathematical formulation of optimal transport. Building on this connection, we then introduce a Wasserstein-based metric that operates on the raw RIR waveforms, exploiting directly the signed temporal structure.

\begin{figure*}[!t]
    \centering
    \subcaptionbox{ESR\label{fig:2Dpos_ESR}}[0.24\textwidth]{%
        \centering
        \includegraphics[width=\linewidth]{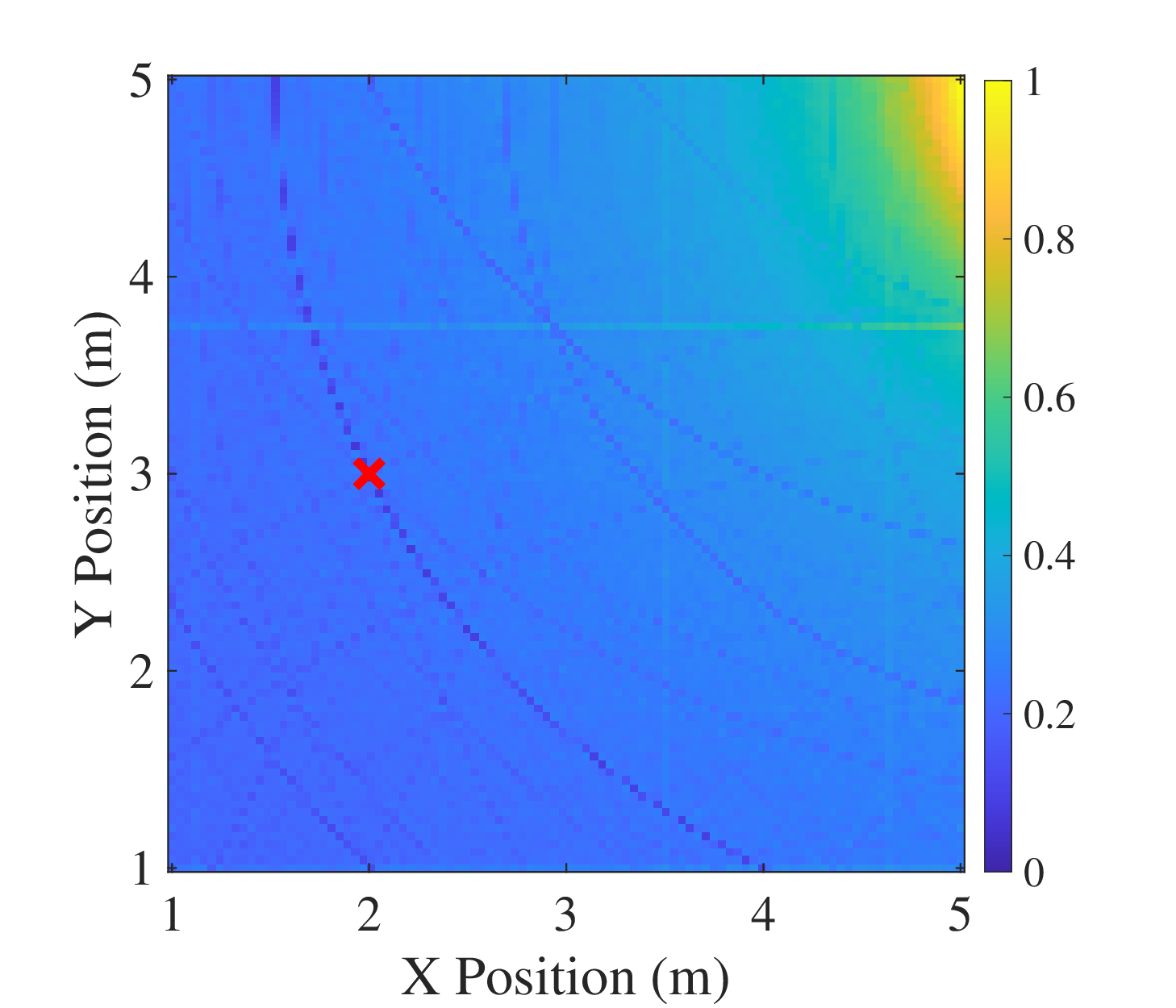}%
    }
    \hfill
    \subcaptionbox{MSS\label{fig:2Dpos_MSS}}[0.24\textwidth]{%
        \centering
        \includegraphics[width=\linewidth]{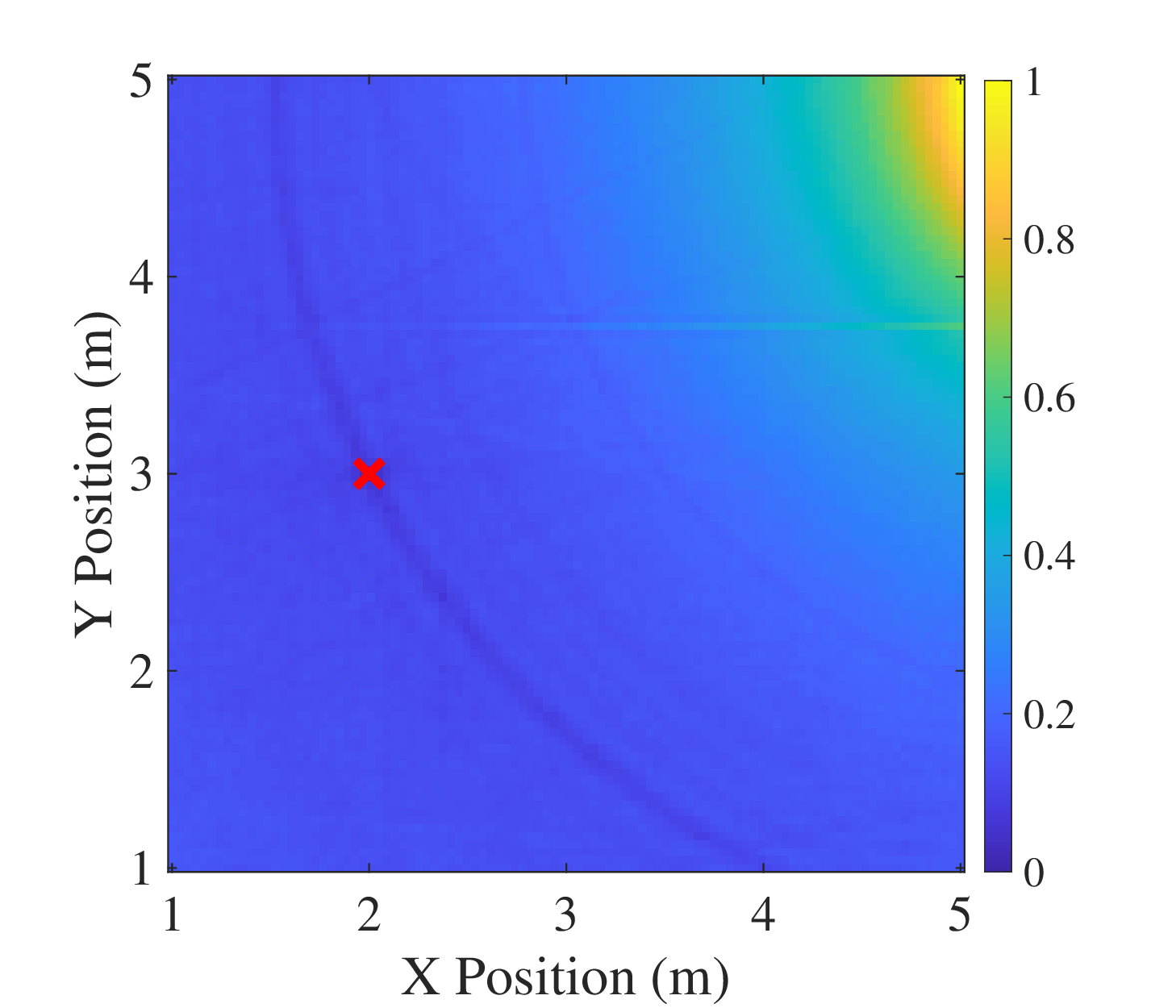}%
    }
    \hfill
    \subcaptionbox{EDC\label{fig:2Dpos_EDC}}[0.24\textwidth]{%
        \centering
        \includegraphics[width=\linewidth]{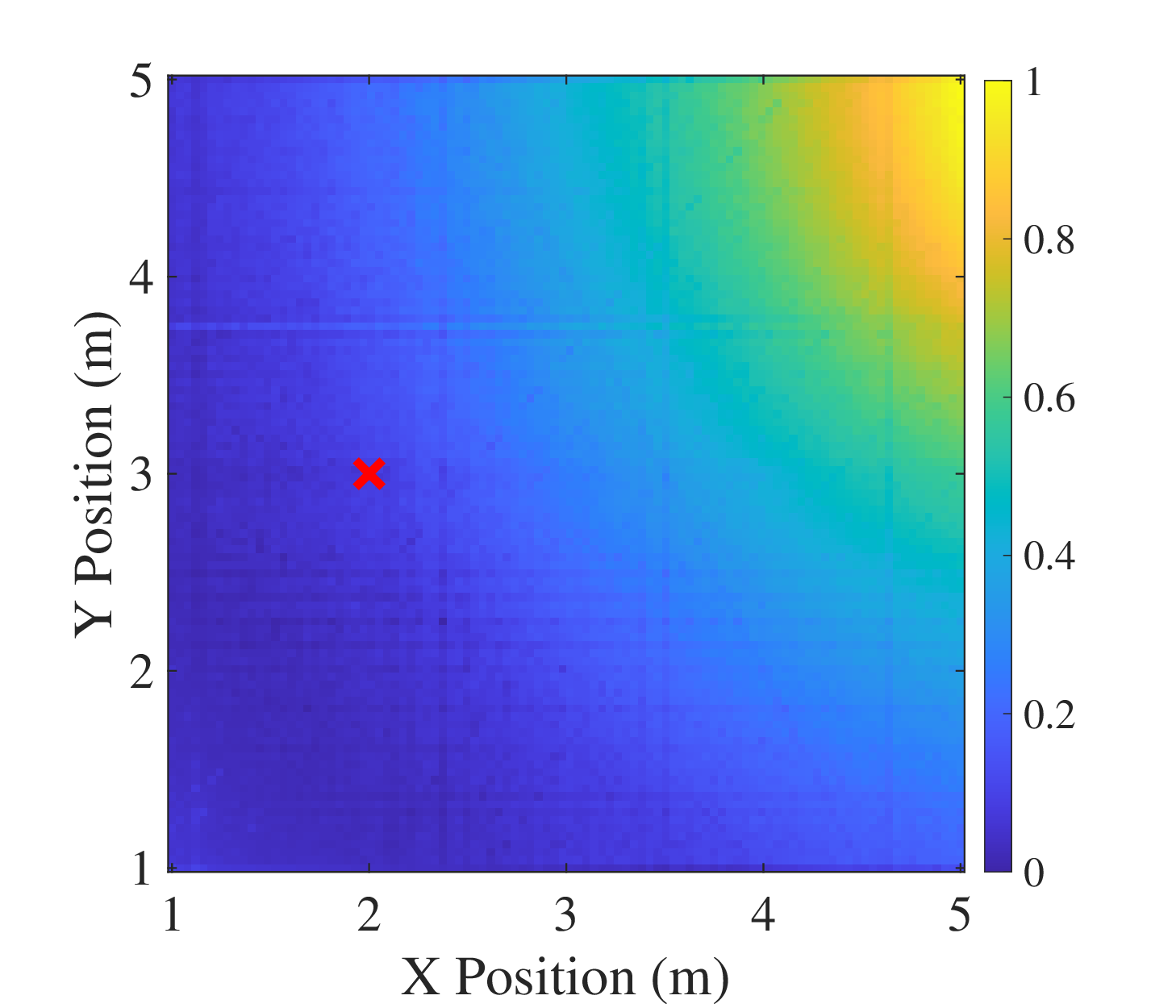}%
    }
    \hfill
    \subcaptionbox{Signed $\mathbb{W}_1$\label{fig:2Dpos_signW1_norm}}[0.24\textwidth]{%
        \centering
        \includegraphics[width=\linewidth]{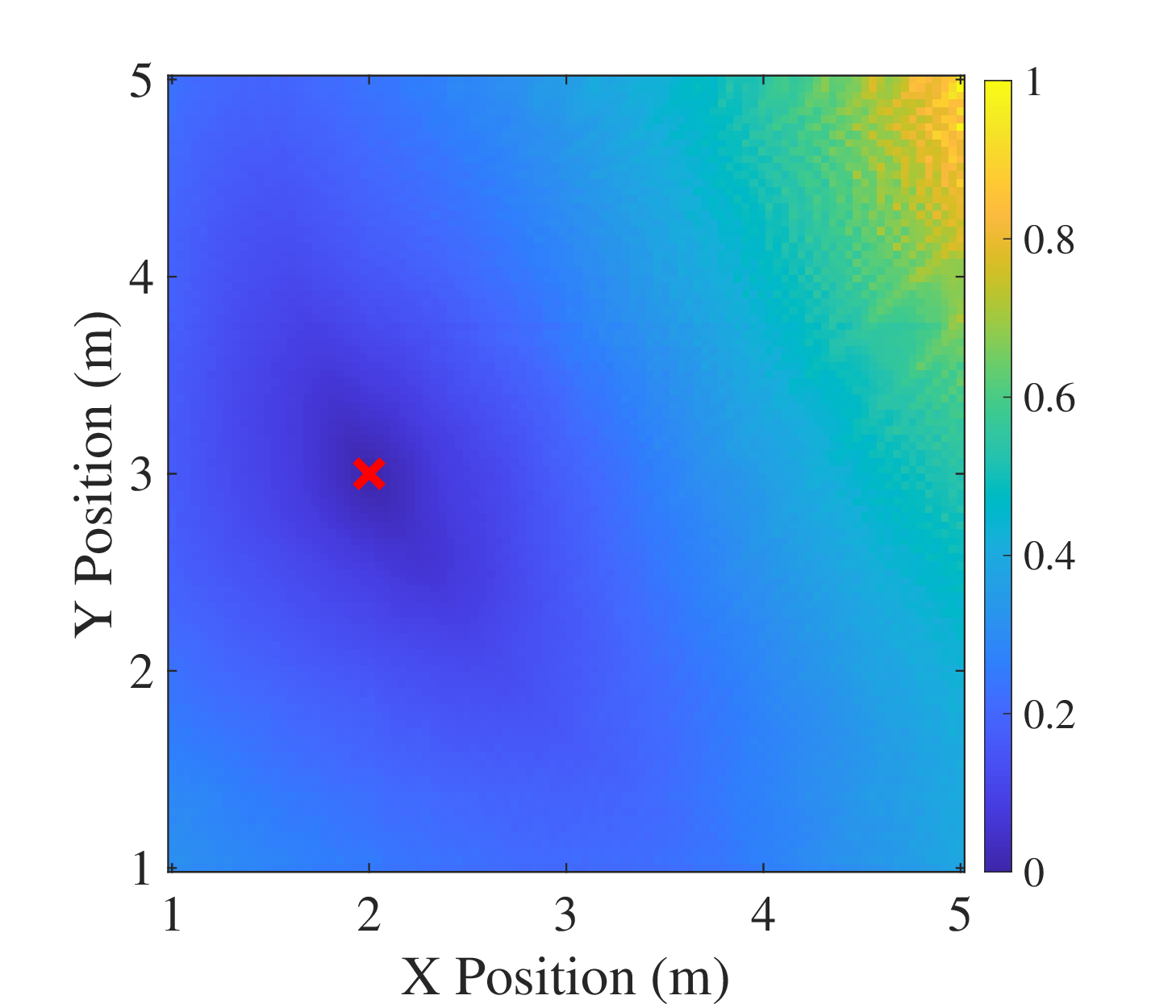}%
    }
    \caption{Two-dimensional source-position robustness of the considered loss functions. The test source was moved over a rectangular grid of positions in $(x,y)\in[1,4]\times[1,4]~\mathrm{m}$ around the target value, while its height was fixed at $z=1.5~\mathrm{m}$.
 All losses are shown on a normalized scale between [0,1].
}
    \label{fig:2Dpos}
\end{figure*}

\subsection{Connection Between EDC and Wasserstein-1}\label{subsec:EDC_W1}

The Energy Decay Curve is a common way to process RIRs, which describes how acoustic energy decays over time inside a room \cite{Schroeder_1965}.
For a RIR $h$ assume that the effective support is limited to the interval $[0,T]$, for $T\in\RR$. Then, the corresponding \textit{linear} EDC, $\edc_h(t)$, is defined through Schroeder’s backward integration as 
\begin{align*}
    %
    %
    \edc_h (t) &\triangleq \int_{\tau = t}^T { dh^2(\tau) } =  \int_{t=0}^T {d h^2(t) } - \int_{\tau=0}^t { dh^2(\tau) }
    \\&= F_{h^2}(T) - F_{h^2}(t),
\end{align*}
where $F_{h^2}$ is the CDF of the squared impulse response $h^2$ \cite{Schroeder_1965}. 

Two impulse responses, $h_1$ and $h_2$, may then be compared in terms of $L_p$ distances between their EDCs. In particular, the $L_p$ distance (to power $p$) is
\begin{align*}
    \Ldist_p(\edc_{h_1},\edc_{h_2})^p &=\!\! \int_{0}^T|\edc_{h_1}(t)-\edc_{h_2}(t)|^pdt \\&=\!\! \int_{0}^T\!\!\!| F_{h_1^2}(t) - F_{h_2^2}(t) - (F_{h_1^2}(T) - F_{h_2^2}(T))|^p.
\end{align*}
Here, by comparing to \eqref{eq:wasserstein1}, it may be noted that if $F_{h_1^2}(T) = F_{h_2^2}(T)$ and one considers $p = 1$, then $\Ldist_1(\edc_{h_1},\edc_{h_2}) = W_1(h_1^2,h_2^2)$. That is, comparing EDCs in terms of $L_1$ is equivalent to comparing the temporal structure of the \textit{energy distributions} $h_1^2,h_2^2\in \posmeas(\RR)$ in terms of the Wasserstein-1 distance, given that the total energy $F_{h^2}(T)$ is the same for both RIRs. This connection shows that EDC comparison already has an implicit transport interpretation: it compares how acoustic energy is distributed over time rather than comparing samples at identical time positions. However, the implication of this is that the energy, as opposed to the amplitudes, of individual RIR peaks is treated as additive. This is not consistent with the geometrical view of room acoustics, where RIRs are the collection of reflections and spatial variations will result in both constructive and destructive interference.

Based on this observation, we propose a similarity metric that shares the attractive property of non-local comparison with the $L_1$ distance for EDCs while also respecting the structured behavior of RIRs. 

\subsection{Signed Wasserstein-1 distance}

The classical Wasserstein distance is defined for non-negative distributions of equal total mass. Since RIRs are generally signed, we extend this transport-based comparison to signed signals using the concept of the \textit{signed Wasserstein distance}. This construction is based on the Jordan decomposition of signed measures into its positive and negative parts, which may then be compared using a non-negative transport problem \cite{mainini2012signed, piccoli2019wasserstein}.
Let $\mu,\nu \in \signedmeas(\RR)$ be two signed measures with equal total mass, i.e., $\mu(\RR) = \nu(\RR)$. It may be noted that $\mu$ and $\nu$ admit Jordan decompositions according to
\begin{equation*}
    \mu = \mu_+ - \mu_- ,  \quad  \nu = \nu_+ - \nu_-,
\end{equation*}
where $\mu_+, \mu_-, \nu_+, \nu_- \in \posmeas(\RR)$. With this, their signed \mbox{Wasserstein-1} distance $\signedwass_1(\mu,\nu)$ is defined as
\begin{align}\label{eq:W1decomp}
    \signedwass_1(\mu,\nu) = W_1(\mu_+ + \nu_-, \mu_- + \nu_+).
\end{align}
In the case of non-negative measures, $\signedwass_1$ and $W_1$ coincide. In our context, we use the model in \eqref{eq:RIR_atomic}, allowing for comparing two RIRs $h_1,h_2 \in\signedmeas(\RR)$ using $\signedwass_1$. However, when modeling RIRs as elements of $\signedmeas(\RR)$, it is not necessarily the case that two RIRs integrate to the same value. Herein, we focus on shape rather than scale properties of RIRs and therefore normalize the considered decomposed positive distributions by their total integral, or, in discrete time, sum. Although this discards absolute gain information, it allows for comparing the temporal structure of RIRs.

\section{Numerical Experiments}
To evaluate the robustness of the proposed RIR metric, we conduct a series of controlled perturbation studies by varying different acoustics parameters. All simulations were performed using synthetic RIRs generated using the Pyroomacoustics library \cite{scheibler2018pyroomacoustics}. Since the simulated RIRs are discrete-time quantities, we use the discrete analogs for all introduced quantities, replacing integrals with sums\footnote{This corresponds to considering interpolated versions of \eqref{eq:RIR_atomic} restricted to a sampling grid \cite{laaksoSplit}.}. For comparison, we provide as baselines the linear EDC formulation from \cref{subsec:EDC_W1}, the Error-to-Signal Ratio (ESR) \cite{Wright_Damskagg_Valimaki_2019}, defined as 

\begin{equation} \mathcal{L}_{\textrm{ESR}}(h, \hat{h}) = \frac{\sum_{t=0}^T | h(t) - \hat{h}(t)|^2}{\sum_{t=0}^T | h(t)|^2}, \end{equation}
and the Multi-Scale Spectral (MSS) loss which combines $\ell_2$ and $\ell_1$ distances magnitude and log-magnitude, respectively, across different Short-Time-Fourier-Transform resolutions \cite{Yamamoto2019ParallelWA, steinmetz2020auraloss}.
\begin{figure}[t]
    \centering
\includegraphics[width=1\columnwidth, trim= 70 0 70 10, clip]{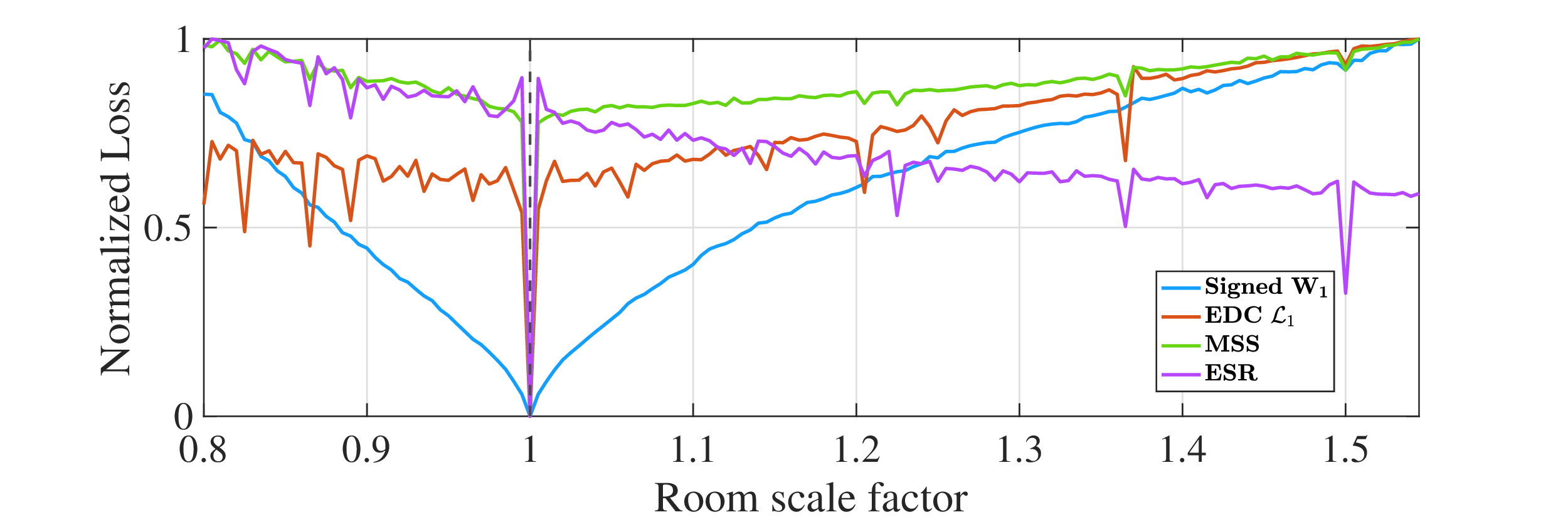}
    \caption{Distance robustness to uniform room dimension scaling along the $(x,y,z)$ axes. All losses are shown on a normalized scale between [0,1].}
    \label{fig:room_scaling}
\end{figure}
\subsection{Source-position shift}\label{subsec:sourcePos}
First, we evaluate the response of the considered distances to perturbations of the sound source position in a room with all other parameters fixed. The experiments were conducted in a shoe-box room of dimensions $(8.0\times6.0\times3.0~\mathrm{m})$, with a reference source at $[2,3,1.5]~\mathrm{m}$, a fixed microphone at $[6,5,1.5]~\mathrm{m}$, and uniform room absorption coefficient $\alpha=0.4$. The sampling frequency was $24$~kHz and the reflection order was set to $N=10$.
\cref{fig:2Dpos} shows the resulting loss landscapes for the different metrics. The signed $\mathbb{W}_1$ metric is the only one that provides a clear and unique minimum region corresponding to the true target location, with a smooth, gradual increase toward the regions of maximum distance. The EDC metric also correctly identifies the minimum area, but provides comparatively little variation over a substantial portion of the displacement range. Similarly for the ESR and MSS, the landscapes are largely uniform, with some circular structures visible. This is likely caused by source positions being equidistant from the microphone, where reflections arrive at the same time instant. Indeed, this experiment is particularly sensitive to the correct evaluation of the direct-path of the sound, which requires precise handling of the early reflections in the RIR. The ability of the Signed $\signedwass_1$ to quantify the horizontal displacement of the peaks then allows for a clear and smooth distance definition. 

\subsection{Room scaling variation}
In the second experiment, we study the robustness of the metrics to variations in the room geometry, by uniformly scaling the room dimensions along the $(x,y,z)$ axes. The same parameters as in ~\cref{subsec:sourcePos} were used, while the three room dimensions are scaled by a common factor over the range $[0.8,1.5]$. In \cref{fig:room_scaling}, one can observe that the MSS and the ESR curves maintain a roughly constant loss level across most of the scaling range, with a sharp decline only in the immediate vicinity of the original room dimension. This suggests that the point-wise comparisons can become relatively insensitive to changes in reflection locations when these undergo temporal shifts. The EDC loss exhibits a similarly narrow minimum around the ground truth room dimensions, although its shape is more symmetric. In contrast, the signed $\mathbb{W}_1$ produces the smoothest and most gradual distance profile across the scaling range. This behavior may be attributed to the fact that the scaling induces an approximately common scaling of the reflection delays, which is naturally captured by the transport formulation, resulting in a gradual and seemingly linear increase in distance.

\subsection{Room parameters inference}
Finally, we present an example application of the proposed metrics in a differentiable acoustic parameter estimation task. Using the differentiable Image-Source-Model framework of \cite{zhi2023differentiable}, room parameters are estimated via gradient-based optimization  with respect to the absorption coefficients $\boldsymbol{\alpha}=[\alpha_1,\ldots,\alpha_N]^\top$ and room dimension parameters $[L_x,L_y,L_z]$. The experiment is repeated 10 times, with the room dimensions and absorption coefficients initialized around their ground-truth values using independent uniform perturbations. The horizontal dimensions $L_x$ and $L_y$ are perturbed by $\pm3~\mathrm{m}$, while the vertical dimension $L_z$ is perturbed by $\pm0.5~\mathrm{m}$. The absorption coefficients are initialized with perturbations of $\pm0.2$. The source and microphone positions were fixed. Table ~\ref{tab:mae_optimization} summarizes results for individual optimization of the two sets of parameters. The signed $\mathbb{W}_1$ loss achieves the lowest Mean Absolute Error (MAE) of the parameters for room dimension recovery, which is consistent with what happens in the room-scaling experiment and indicates that the transport-based formulation is particularly effective at explaining the temporal shifts of reflections. While all absorption coefficient recovery results are reasonably good, ESR achieves the lowest absorption MAE, suggesting that the metrics exhibit similar sensitivities to amplitude-related changes. The present experiments optimize geometry and absorption coefficients separately; joint optimization represents a more challenging setting due to the coupling between these parameters and is therefore left for future investigation.

\begin{table}[!t]
\caption{Room dimensions and absorption coefficients parameters optimization results}
\label{tab:mae_optimization}
\centering
\small
\begin{tabular}{lcc}
\textbf{Loss} & \textbf{Dimension MAE} & \textbf{Absorption MAE} \\
\midrule
ESR
          & 2.5794 $\pm$ 2.4296 &  0.0004 $\pm$ 0.0001  \\
MSS
          & 1.2276 $\pm$ 1.0114 & 0.0017 $\pm$ 0.0008 \\
EDC
          & 1.5995 $\pm$ 0.8401 & 0.0096 $\pm$ 0.0054 \\
Signed $\mathbb{W}_1$
          & \textbf{0.4208 $\pm$ 0.4963} & 0.0115 $\pm$ 0.0070 \\
\bottomrule
\end{tabular}
\end{table}

\section{Conclusion}
This paper presents a novel similarity metric for comparing raw impulse response signals. The proposed method is based on the theory of signed Wasserstein distances, and allows us to evaluate the similarity directly using the signed reflections of a RIR, which would not be possible using a conventional Wasserstein-1 distance formulation. We demonstrate the robustness of the proposed metric against commonly used point-wise distances in audio processing. The results emphasize the advantageous loss landscape of our signed Wasserstein metric in a series of perturbation studies on room dimensions and source localization. Additionally, the distance is used as a loss function for absorption coefficients and room dimension parameter optimization in a differentiable Image-Source-Model, showing promising results for the latter task. 




\bibliographystyle{IEEEbib}
\bibliography{strings,refs}

\end{document}